\documentstyle[aps,prl,multicol,epsf]{revtex}

\begin{document}

\title{Condensate Fragmentation in a New Exactly Solvable
 Model for Confined Bosons}
\author{J. Dukelsky$^1$ and P. Schuck$^2$}
\address{
$^{1}$Instituto de Estructura de la Materia, C.S.I.C.,Madrid, Spain.\\
$^{2}$Institut de Physique Nucl\'eaire, Universit\'e Paris-Sud, France.}

\maketitle

\begin{abstract}
Based on Richardson's exact solution of the multi level pairing model we derive a new class of exactly solvable models for
finite boson system. As an example we solve a particular hamiltonian which
displays a transition to a fragmented condensate for repulsive pairing interaction.

\noindent
PACS number: 03.75.Fi, 05.30.Jp

\end{abstract}

\begin{multicols}{2}

Since the experimental realization of condensates of trapped bosonic atoms
there exists a considerable growing interest in refining the approximations
to study finite many body boson systems. Standard approaches used up to now
are mean field based theories (MFT) like the
Gross-Pitaevskii equation and the Bogoliubov theory which considers pair
fluctuations on top of a Bose condensate. In order to test such many body approximations and also to gain insight, exactly solvable models are extremely useful. Two of such exactly solvable models were used in the context of boson traps, the one dimensional hard core boson model \cite
{Girar} and the harmonic interaction model \cite{Gaj}. Both model studies
show the need to go beyond MFT. We would like to present in this letter a
new class of exactly solvable models based on a generalization of Richardson's exact solution of the pairing model\cite{richa0,richa1,richa2}. Of particular interest will be the model case with repulsive pairing interaction, where we find a fragmented condensate with two macroscopic eigenvalues of the one body density matrix corresponding to the lowest $s$ and $p$ wave states for bosons in a confining potential. The ocurrence of condensate fragmentation in confined boson systems is a much debated subject at present and so far no clear evidence for the existence of such a state has been demonstrated. Nozieres and Saint-James \cite{Noz} using the Hartree-Fock Bogolibov (HFB) approximation and later on Nozieres \cite {Noz2} using Hartree-Fock (HF) theory showed that fragmentation is energetically unfavoured with respect to a single condensate state for scalar boson systems. Going beyond mean field  Holzmann $et. al.$ \cite {Holz,Mu} discussed the possible existence a fragmented scalar boson condensate in two dimensions.
The case of trapped spin 1 boson systems has been treated in refs \cite {Law,Ho} where it was concluded that fragmentation might be realized in spin space but they still used mean field theory to describe the condensates in each spin component.         

Let us now come to the construction of our new exactly solvable model for interacing bosons.     
In ref.\cite{richb} Richardson was able to obtain the exact eigenstates of a
pairing hamiltonian for bosons

\begin{equation}
H_{P}=\sum_{\Lambda }\varepsilon _{\Lambda }~n_{\Lambda }+\frac{g}{2}\sum_{\Lambda
\Lambda ^{\prime }}A_{\Lambda }^{\dagger }A_{\Lambda ^{\prime }}  \label{H1}
\end{equation}

where $A_{\Lambda }=\sum_{\alpha }$ $a_{\Lambda \overline{\alpha }%
}a_{\Lambda \alpha \text{ }}$, $n_{\Lambda }=\sum_{\alpha }a_{\Lambda \alpha
}^{\dagger }a_{\Lambda \alpha }$ , $\left( \Lambda \alpha \right) $ are the
quantum numbers of a confining potential (in the particular example of a three dimensional spherical potential which we will discuss below $\Lambda $ is the principal quantum number and $\alpha$ represents the orbital quantum numbers), and $\left( \Lambda \overline{\alpha} \right) $ is the time reversal of $\left( \Lambda \alpha \right) $.   The operator $A_{\Lambda }^{\dagger }$ creates a zero angular momentum pair (singlet) in level $\Lambda$ and the operator $n_{\Lambda }$ counts the number of bosons in level $\Lambda$. 

In analogy with the solution for fermions Richardson
assumed that the hamiltonian eigenstates can be written as a pair product
wave function of the form

\begin{equation}
\left| \psi \right\rangle =\prod_{i=1}^{{\cal N}}B_{i}^{\dagger }\left|
\varphi \right\rangle  \label{wave}
\end{equation}
where $B_{i}^{\dagger }=\sum_{\Lambda }\frac{1}{2\varepsilon_{\Lambda
}-E_{i}}A_{\Lambda }^{\dagger }$ , ${\cal N}$ is the number of singlet boson pairs, $%
E_{i}$ are pair energies to be determined by the eigenvalue equation and $%
\left| \varphi \right\rangle $ is a state of $\nu$ unpaired bosons that fulfills $%
A_{\Lambda }\left| \varphi \right\rangle =0$ and $n_{\Lambda }\left| \varphi
\right\rangle =\nu _{\Lambda }\left| \varphi \right\rangle $. A given configuration in the Hilbert space is determined by the set ${\nu _{\Lambda }}$ of unpaired bosons in each level $\Lambda$. The total number of bosons is then $ N = 2 {\cal N}+ \sum_{\Lambda }\nu _{\Lambda }$.
A crucial point is the observation that the off diagonal matrix elements of the one body density matrix $\left\langle \psi \right| a_{\Lambda \alpha }^{\dagger }a_{\Lambda ^{\prime
}\alpha ^{\prime }}\left| \psi \right\rangle$ are zero with $\left| \psi \right\rangle$ given by (\ref{wave}) because the action of $a_{\Lambda \alpha }^{\dagger }a_{\Lambda ^{\prime
}\alpha ^{\prime }}$ on $\left| \psi \right\rangle$ either destroys a singlet pair increasing $\nu$ by $2$ or it acts on an unpaired particle chanching its configuration from $\left( \Lambda \alpha \right) $ to $\left( \Lambda^{\prime} \alpha^{\prime} \right) $. In any case the state 
$a_{\Lambda \alpha }^{\dagger }a_{\Lambda ^{\prime}\alpha ^{\prime }}\left| \psi \right\rangle$
with $\Lambda \alpha \ne \Lambda^{\prime } \alpha^{\prime }$ will be orthogonal to $\left\langle \psi \right|$. Therefore the occupation numbers $\left\langle \psi \right| n_{\Lambda} \left| \psi \right\rangle$ are the eigenvalues of the one body density matrix. This situation is analogous to the Fermion case where this is known as the generalised seniority scheme \cite{talmi}.
     
By acting with the hamiltonian on the trial wavefunction, after a straightforward but long
derivation, Richardson obtained the following set of equation for the pair energies

\begin{equation}
1+g\sum_{\Lambda}\frac{\left( \Omega _{\Lambda }+2\nu _{\Lambda }\right) }{\left(
2\varepsilon_{\Lambda }-E_{i}\right) }+4g\sum_{j\neq i}\frac{1}{E_{j}-E_{i}}=0
\label{richar}
\end{equation}
where $\Omega _{\Lambda }$ is the degeneracy of level $\Lambda $ .The
energy eigenvalues of (\ref{H1}) are given by $E=\sum_{\Lambda }\varepsilon _{\Lambda }\nu
_{\Lambda }+\sum_{i=1}^{{\cal N}}E_{i}$ for each solution of the set of
equations (\ref{richar}) as will be explained below.

The proof of integrability of the pairing model for fermion systems has been given by Cambiaggio $et$ $al.$ \cite
{Sarra}. They found a complete set of commuting operators (constants of motion)in terms of which
it is possible to express the pairing hamiltonian and the number
operator as particular linear combinations. Unfortunately, since the authors
were not aware of the Richardson's previous works, they were not able to obtain the
eigenvalues of the new set of operators, nor the exact solution of the
pairing hamiltonian. This connection has been established recently \cite
{sie2} and the eigenvalues of the commuting operators found using Conformal
Field Theory.

The Richardson solution for
fermions remained practically unused till very
recently when it was applied, in the context of ultrasmall superconducting
grains \cite{sie1}, to study the transition from the superconducting to the normal state. We are
not aware of any application for boson systems.

In complete analogy to the fermion case, a set of global commuting operators can also be written in the boson case,
based on the group
algebra of the pseudo-spin generators $K_{\Lambda }^{0}=\frac{1}{2}n_{\Lambda
}+\frac{1}{4}\Omega _{\Lambda }$ , $K_{\Lambda }^{+}=\frac{1}{2}A_{\Lambda
}^{\dagger }=\left( K_{\Lambda }^{-}\right) ^{\dagger }$of $SU(1,1)$ .
\begin{eqnarray}
R_{\Lambda }=K_{\Lambda }^{0}+2g\sum_{\Lambda ^{\prime }\left( \neq \Lambda
\right) }\frac{1}{\eta _{\Lambda }-\eta _{\Lambda ^{\prime }}}~K_{\Lambda
}\cdot K_{\Lambda ^{\prime }} \label{ope}
\end{eqnarray}

The scalar product in ( \ref{ope}) refers to the $SU(1,1)$ group, $K_{\Lambda
}\cdot K_{\Lambda ^{\prime }}= \frac{1}{2}\left( K_{\Lambda }^{+}K_{\Lambda ^{\prime }}^{-}+K_{\Lambda
}^{-}K_{\Lambda ^{\prime }}^{+}\right) -K_{\Lambda }^{0}K_{\Lambda ^{\prime
}}^{0}$.
The set of operators $R_\Lambda$ is complete, the operators commute among each other, and the pairing hamiltonian (\ref{H1}) can be written as the linear combination $2\sum_{\Lambda }\eta _{\Lambda } R_{\Lambda }$. These conditions demonstrate that the pairing model is integrable for boson systems.  

 We will now look for the eigenstates of
the $R_{\Lambda }$ operators (\ref{ope}) using the trial eigenstates (\ref
{wave}). The form of the boson pair amplitudes $u_{\Lambda }^{i}$ in the
boson pair operators $B_{i}=\sum_{\Lambda }u_{\Lambda }^{i}A_{\Lambda }$ is
fixed by the solution of the one pair problem, namely $u_{\Lambda}^{i}=1/\left(
2\eta _{\Lambda}-E_{i}\right) $. The eigenvalues $R_{\Lambda }\left| \psi
\right\rangle =\lambda _{\Lambda }\left| \psi \right\rangle $ can be worked
out in an analogous way as in the original Richardson paper. The eigenvalues
are given by

\begin{equation}
\lambda _{\Lambda }=\frac{\Sigma _{\Lambda }}{2}\left[ \frac{1}{2}-2g\sum_{i}%
\frac{1}{2\eta _{\Lambda }-E_{i}}-\frac{g}{4}\sum_{\Lambda ^{\prime }\neq
\Lambda }\frac{\Sigma _{\Lambda ^{\prime }}}{\eta _{\Lambda }-\eta _{\Lambda
^{\prime }}}\right]  \label{a1}
\end{equation}
where $\Sigma _{\Lambda }=\Omega _{\Lambda }+2\nu _{\Lambda }$, while the
pair energies $E_{i}$ should be the roots of the Richardson equations (\ref
{richar}) with $\eta_\Lambda$ replacing $\varepsilon_\Lambda$. 

One can readily verify that for $\eta_\Lambda=\varepsilon_\Lambda$ in eqs. (\ref{ope}) and (\ref{a1}), the linear combination $2\sum_{\Lambda }\varepsilon _{\Lambda }R_{\Lambda }$
gives the Richardson hamiltonian (\ref{H1}) and $2\sum_{\Lambda} \varepsilon _{\Lambda } 
\lambda _{\Lambda }$ gives
the corresponding eigenvalue. An important by-product of having found the
eigenvalues (\ref{a1}) is that the $\eta _{\Lambda}$ are in principle free
parameters not necessarily related to the single boson energies. We will
exploit this freedom to obtain solutions for generalized pairing
hamiltonians.

Looking back to the form of the hamiltonian (\ref{H1}) one sees that since the pair operators $A_{\Lambda }$ are normalized to the square root of the degeneracy of the level $\Lambda$, the
effective pairing matrix elements are proportional to the square root of the product of the degeneracies of the two shells $%
\sqrt{\Omega _{\Lambda }\Omega _{\Lambda ^{\prime }}}$. In a spherical harmonic
oscillator potential each degeneracy is in turn proportional to $\Lambda^{d-1}$
where $d$ is the space dimension and $\Lambda$ is the principal quantum
number while the single boson energies are linear in $\Lambda$, producing unphysical occupations of the high lying levels for
attractive pairing or a compression of the bosons in the lowest level for
repulsive pairing. We will obtain a more realistic hamiltonian making use
of the freedom in the choice of the $\eta _{\Lambda}$. In order to cancel the undesired dependence of the effective pairing matrix elements on the degeneracies we make the following definition $\eta_\Lambda = (\varepsilon_\Lambda )^{d}$ in eqs. (\ref{ope}) and (\ref{a1}). The new hamiltonian, $ H=2 \sum_{\Lambda} \varepsilon_{\Lambda} R_{\Lambda}$, can be expanded using the definition of the $R_\Lambda$ (\ref{ope}) in terms of the $SU(1,1)$ generators or equivalently in terms of the pair operators.
The final form of the hamiltonian is (more details of its construction will be given elsewhere)

\begin{equation}
H={\cal E}+\sum_{\Lambda}\overline{\varepsilon }_{\Lambda}\ n_{\Lambda}+
\sum_{\Lambda \Lambda^{\prime }} V_{\Lambda \Lambda^{\prime }} {\left( A_{\Lambda}^{\dagger }A_{\Lambda^{\prime }}-n_{\Lambda}n_{\Lambda^{\prime
}}\right)}  \label{hh} 
\end{equation}

with $\cal E$ an uninteresting constant, $\overline{\varepsilon }_{\Lambda}=\varepsilon _{\Lambda}+2 V_{\Lambda \Lambda}-\sum_{\Lambda^{\prime }}V_{\Lambda \Lambda^{\prime }}$ $\Omega _{\Lambda^{\prime }}$ being the single boson energies, and $V_{\Lambda \Lambda^{\prime }}=g/{2 \sum_{l=0}^{d-1}\varepsilon _{\Lambda}^{l}\varepsilon_{\Lambda^{\prime }}^{d-l-1}}$.
 
We can readily check that now the effective interaction terms in (\ref{hh}), contrary to (\ref{H1}), correctly scale with energy. In addition to the pairing term in (\ref{hh}) also appears a particle-hole interaction of the monopole-monopole type which gives the hamiltonian a rather rich and quite general character. It has, however, the restriction that pairing and particle-hole interactions are linked to be of opposite sign, a feature which may be realised only in particular situations. On the other hand, we think,that the equality in magnitude of both interactions do not invalidate our general conclusions below.

The energy eigenvalues of (\ref{hh}) can be obtained summing the eigenvalues (\ref{a1})
as

\begin{equation}
E=\frac{1}{2}\sum_{\Lambda}\varepsilon _{\Lambda}\Sigma _{\Lambda}-\frac{1}{4}\sum_{\Lambda\neq
\Lambda^{\prime }} \Sigma _{\Lambda}\Sigma _{\Lambda^{\prime }} V _{\Lambda \Lambda^{\prime }}%
-2g\sum_{\Lambda i}\frac{\varepsilon _{\Lambda}\Sigma _{\Lambda}}{2\eta _{\Lambda}-E_{i}}
\label{ener}
\end{equation}

It is worthwhile to emphasize at this point that the states (\ref{wave}) are
common eigenstates of the operators $R$ in (\ref{ope}) and, consequently, of any linear
combination of them like the pairing hamiltonian (\ref{H1}) or our more
general hamiltonian (\ref{hh}) provided that the pair energies are the solutions
of equation (\ref{richar}). The solution of (\ref{richar}) was already
discussed in ref.\cite{richb} and we will give here only a brief summary.
Assuming that we have $L$ oscillator levels and $2{\cal N}$ paired bosons, the ${\cal N}$ pair energies $E_{i}$ are
real roots of (\ref{richar}) in the interval $-\infty <E_{i}<2 \eta _{L}$ for
an attractive pairing interaction or $2 \eta _{0}<E_{i}<\infty $ for a repulsive
pairing interaction. The groundstate has all pair energies in the restricted
intervals $-\infty <E_{i}<2 \eta _{0}$ for an attractive pairing interaction or $2 \eta
_{0}<E_{i}<2 \eta _{1}$ for a repulsive pairing interaction. Any state in the Hilbert
space corresponds to a particular distribution of the ${\cal N}$ pair
energies into the $L$ intervals. States with broken pairs can be generated
by replacing a boson pair $B$ by two unpaired bosons which can occupy
any of the $L$ shells. For example, the first excited state with two
unpaired bosons corresponds to solve (\ref{richar}) with ${\cal N}-1$ pair
energies in the first possible interval and $\nu _{0}=\nu _{1}=1.$

Having in mind that the eigenstates of the hamiltonian (\ref{hh}) are the
same as those of the pairing hamiltonian (\ref{H1}), the occupation numbers
for a given state can be calculated as the derivatives of the pairing
hamiltonian $H_{P}$ (\ref{H1}) with respect to the the single boson energies $\varepsilon$ as
has been done by Richardson in \cite{richb}

\begin{equation}
\left\langle n_{\Lambda}\right\rangle =\left\langle \frac{\partial H_{P}}{\partial
\varepsilon _{\Lambda}}\right\rangle =\nu _{\Lambda}+\sum_{i}\frac{\partial E_{i}}{\partial
\varepsilon _{\Lambda}}  \label{oc}
\end{equation}

Details of the derivation of the occupation numbers and the final set of
equations can be found in ref.\cite{richb}.

\begin{figure}
\hspace{0.5cm}
\epsfysize=8cm
\epsfxsize=7cm
\epsffile{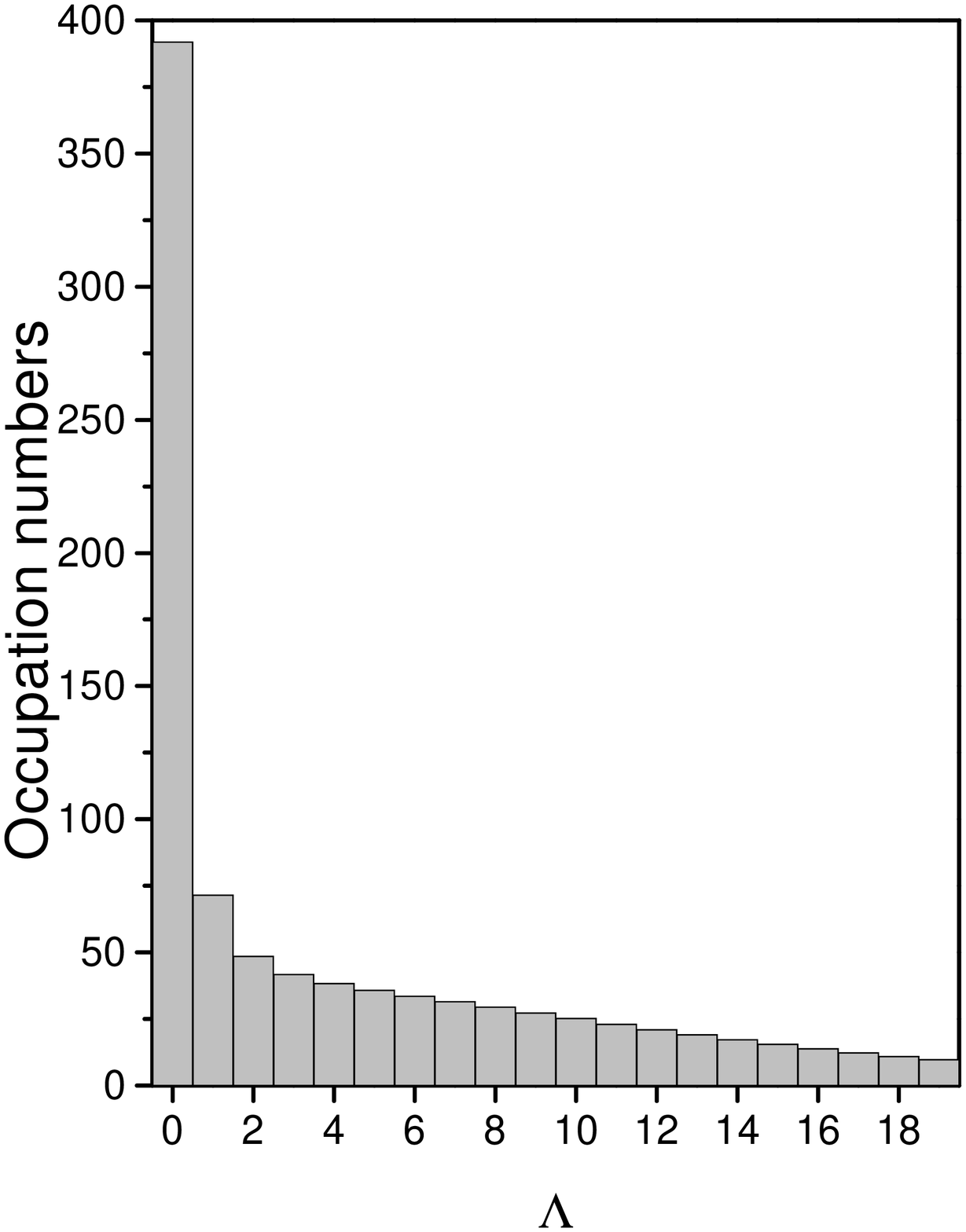}
\narrowtext
\caption{Occupation numbers for $1000$ bosons in $50$ harmonic oscillator
shells, the interaction strength is $g=-2.0$, giving a depletion factor of 0.608. }
\label{fig1}
\end{figure}

We have performed a series of calculations for a system of $1000$ bosons $%
\left( {\cal N}=500\right) $ trapped in a three dimensional spherical
harmonic oscillator with a cutoff at $101/2\ \hbar \omega $ ($L=50)$. For
attractive pairing all quantities evolve smoothly with increasing
attraction. In fig. 1 we show the occupation numbers for $g=-2.0$
corresponding to a depletion factor of $0.608$. They display a reasonable
pattern filling first the lowest levels. This is not the case for the
pairing hamiltonian (\ref{H1}) with $\eta_\Lambda=\varepsilon_\Lambda$
 which for $%
d>1$ populates first the high lying levels. The comparison between the exact results and approximations like Hartree-Fock-Bogoliubov or their number conserving extensions
\cite{giar,gar} will
be the topic of a future work\cite{duss}. 

Changing now to repulsive pairing we have found an unexpected feature. For
some critical value of $g$ the normal groundstate boson condensate suddenly
turns into a state in which the bosons are condensed into the $%
\Lambda=0 $ and the $\Lambda=1$ states while the occupation of the other levels is
negligible. This state was already envisaged by Richardson
himself studying an approximate solution for eq. (\ref{richar}) in the
thermodynamic limit \cite{richb}. Taking into account that the one body density matrix is diagonal in the basis $(\Lambda \alpha)$ (see above) the occupation numbers in (\ref{oc}) are their diagonal matrix elements or, equivalently, the density matrix eigenvalues. Therefore, this new state with a macroscopic occupation of the two lowest harmonic oscillator shells constitutes a truly fragmented condensate state.    
It is commonly accepted since the work of Nozieres and Saint James \cite{Noz} that
for homogeneous systems fragmentation cannot occur in systems of scalar bosons with repulsive interactions. This might perhaps be the first example of fragmentation in a confined boson systems.

In order to understand the nature of this new quantum phase we consider a coherent state $%
\left| \phi \right\rangle =\exp \left[ \sqrt{2{\cal N}}\Gamma ^{\dagger
}\left| 0\right\rangle \right] $ where $\Gamma $ is the most general time
reversal invariant coherent boson that breaks rotational and
reflection symmetry.

\[
\Gamma ^{\dagger }=\frac{1}{\sqrt{1+\beta ^{2}}}\left[ a_{0}^{\dagger
}+\beta \cos \gamma \ a_{110}^{\dagger }+\right.
\]

\begin{equation}
\left. \frac{\beta \sin \gamma }{\sqrt{2}}\left( e^{i\varphi
}a_{11-1}^{\dagger }-e^{-i\varphi }a_{111}^{\dagger }\right) \right]
\label{gama}
\end{equation}

where the three independent variables are defined in the intervals $\beta
\geq 0,\ 0\leq \gamma \leq \pi $ and 0$\leq \varphi \leq 2\pi $.

Since, as mentioned before, the groundstate is a common eigenstate to our
hamiltonian (\ref{hh}) 
 and to the pairing hamiltonian (\ref{H1}) with $\eta _{\Lambda }=\left( \hbar \omega \Lambda +3/2\right) ^{3}$, for
simplicity we minimize the energy of the latter hamiltonian which, indeed, 
will turn out to be a very accurate procedure. Apart from a
constant energy term, the groundstate energy and the occupation numbers are
given by

\begin{equation}
E=\frac{\beta ^{2}}{1+\beta ^{2}}+\frac{x\left( 1-\beta ^{2}\right) ^{2}}{%
4\left( 1+\beta ^{2}\right) ^{2}}  \label{ecohe}
\end{equation}

\begin{equation}
n_{0}=\frac{2{\cal N}}{1+\beta ^{2}}\ ,\ n_{1}=\frac{2{\cal N}\beta ^{2}}{%
1+\beta ^{2}}  \label{oc2}
\end{equation}

where we have defined the adimensional parameter $x=\frac{4{\cal N}g}{\hbar
\omega }$. The energy (\ref{ecohe}) is independent of the variational
parameters $\gamma $ and $\varphi $. Minimization of (\ref{ecohe}) with
respect to $\beta $ gives the solutions $\beta =0$ for $x\leq 1$ and $\beta =%
\sqrt{\frac{x-1}{x+1}}$ for $x>1$. The critical interaction strength
corresponding to the phase transition at $x=1$ is $g_{c}=\frac{\hbar \omega
}{4{\cal N}}$. On the other hand, it is easy to check that the Bogoliubov
approximation has a break down at this critical value of $g$ showing the instability of the singlet boson condensate against pair fluctuations.
Beyond the critical point the occupation numbers for the first two
levels, obtained by inserting the minimum $\beta $ value in (\ref{oc2}),
 are $n_{0}=\frac{x+1}{2x}$ and $n_{1}=\frac{x-1}{2x}$.

\begin{figure}
\hspace{0.6cm}
\epsfysize=7cm
\epsfxsize=6cm
\epsffile{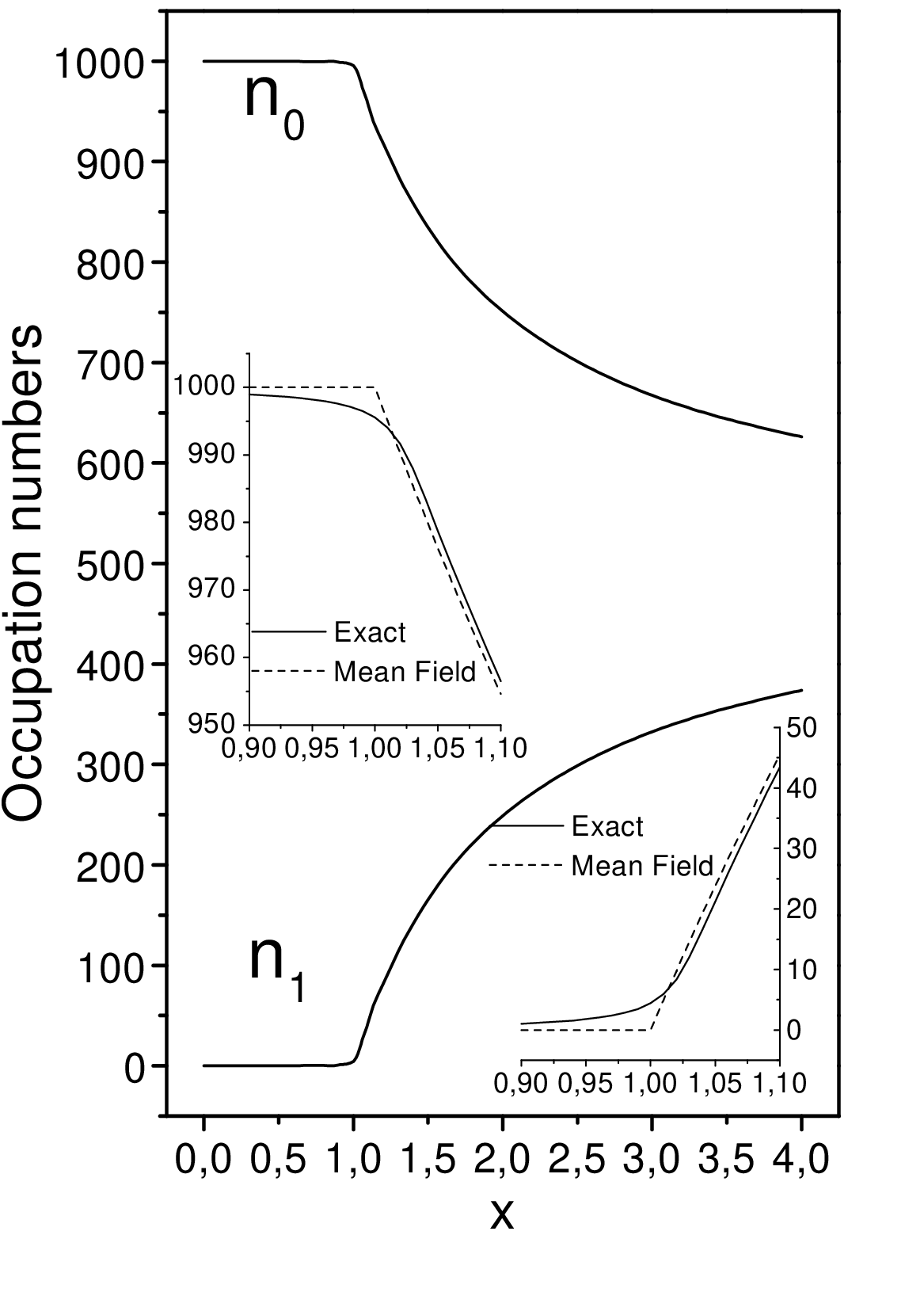}
\caption{Occupation numbers $n_0$ and $n_1$ for $1000$ bosons in $50$
harmonic oscillator shells as a function of the adimensional parameter $x$.}
\narrowtext

\label{fig2}
\end{figure}

In figure 2 we have plotted the occupation numbers of the first two levels
for a system of $1000$ bosons interacting in $50$ levels of a three
dimensional harmonic oscillator as a function of the parameter $x$ . The
exact results in solid lines clearly display an abrupt transition for $x=1$
as predicted by the meanfield description. In fact the approximate results
are indistinguishable from exact ones. Moreover the total occupation of the
rest of the levels is always lower than $10^{-2}$. We have added two insets
to the figure to show a close up of the transition region for the
occupations $n_{0}$ and $n_{1}$. The small differences between the exact and
the approximate results suggest that this is a $1/{\cal N}$ effect, and that
we are seeing the precursor of a true quantum phase transition.

The new phase is a rather peculiar fragmented state characterized by a
macroscopic occupation of only the two lower levels. It can be approximated by a single condensate state at the price of breaking of the reflection symmetry generating a permanent
dipole deformed state.
As we have seen before the parameter $x$ is
proportional to the number of bosons and inversely proportional to the
oscillator frecuency. Assuming that this phase
transition might be realized in realistic systems, it can be controlled by varying
the number of bosons or the characteristics of the confining potential.

In conclusion, we have developed a new class of exactly solvable models for
confined boson systems. These models are exactly solvable in any dimension (reflected in the degeneracies $\Omega_\Lambda$) and with any kind of confining potential (reflected in the single boson energies $\varepsilon_\Lambda$). We have made numerical applications showing the great
potential utility of the model for serving as a testing tool for many body
approximations. Moreover we have found a quantum phase transition to a fragmented state for repulsive
pairing interactions. To the best of our knowledge, this is the first example of fragmentation in a confined scalar boson system. As such, it may stimulate further investigations to see whether this new phase can arise in more realistic situations.

{\bf Acknowledgments} We thank G. Sierra and G.G. Dussel for fruitful conversations. This
work was supported in part by the DGES spanish grant PB98-0685.

\end{multicols}

\end{document}